\documentclass[aps,pra,twocolumn,groupedaddress]{revtex4-1}

\usepackage[table]{xcolor}
\usepackage{colortbl}
\usepackage{indentfirst,avant,tabularx}

\usepackage{amssymb,amsmath,amsthm,amsfonts,amsbsy}
\usepackage{bm,bibunits,color,chngcntr,epsfig,epstopdf,graphicx,dsfont,hyperref,lipsum,,makecell,mathrsfs,rotating}
\usepackage[english]{babel}
\usepackage[normalem]{ulem}
\usepackage[utf8]{inputenc}
\usepackage[skins]{tcolorbox}

\newcommand{\be}{\begin{equation}}\newcommand{\ee}{\end{equation}}
\newcommand{\bea}{\begin{eqnarray}}\newcommand{\eea}{\end{eqnarray}}
\newcommand{\ket}{\rangle}\newcommand{\bra}{\langle}

\newcommand{\ra}{\rightarrow}

\newcommand{\I}{\mathds{1}}
\def\C#1{\mathcal #1}

\def\ps{|\psi\ket}

\begin{document}
\newtheorem{theorem}{Theorem}
\newtheorem{proposition}[theorem]{Proposition}
\newtheorem{corollary}[theorem]{Corollary}
\newtheorem{open problem}[theorem]{Open Problem}
\newtheorem{definition}{Definition}
\newtheorem{remark}{Remark}
\newtheorem{example}{Example}
\newtheorem{Question}{Question}

\title{A prototype of quantum von Neumann architecture}

\author{Dong-Sheng Wang}
\affiliation{CAS Key Laboratory of Theoretical Physics, Institute of Theoretical Physics,
Chinese Academy of Sciences, Beijing 100190, China}



\begin{abstract}
A modern computer system, based on the von Neumann architecture, 
is a complicated system with several interactive modular parts. 
It requires a thorough understanding of the physics of information
storage, processing, protection, readout, etc.
Quantum computing, as the most generic usage of quantum information,
follows a hybrid architecture so far, namely,
quantum algorithms are stored and controlled classically,
and mainly the executions of them are quantum,
leading to the so-called quantum processing units.  
Such a quantum-classical hybrid is constrained by its classical ingredients,
and cannot reveal the computational power of a fully quantum computer system 
as conceived from the beginning of the field.
Recently, the nature of quantum information has been further recognized, 
such as the no-programming and no-control theorems,
and the unifying understandings of quantum algorithms and computing models.
As a result, in this work we propose a model of universal quantum computer system,
the quantum version of the von Neumann architecture.
It uses ebits (i.e., Bell states) as elements of the quantum memory unit,
and qubits as elements of the quantum control unit and processing unit.
As a digital quantum system,
its global configurations can be viewed as tensor-network states. 
Its universality is proved by the capability to execute quantum algorithms
based on a program composition scheme via a universal quantum gate teleportation.
It is also protected by the uncertainty principle, 
the fundamental law of quantum information, 
making it quantum-secure distinct from the classical case.
In particular, 
we introduce a few variants of quantum circuits,
including the tailed, nested, and topological ones,
to characterize the roles of quantum memory and control,
which could also be of independent interest in other contexts. 
In all, our primary study
demonstrates the manifold power of quantum information and paves the way 
for the creation of quantum computer systems in the near future.\\

\end{abstract}

\maketitle

\section{Introduction}\label{sec:int}


In quantum computing, we often apply sequences of unitary operations on multi-qubit states,
followed by measurements. 
This is described in the quantum circuit model, the most popular model for universal quantum computing~\cite{NC00}.
Being universal is not only vital to prove its own consistency~\cite{Deu85,BV97,Yao93},
but also to demonstrate its power relative to conventional computing.
However, the quantum circuit model is not complete so far in the sense that 
it does not provide a \emph{quantum computer system},
the quantum version of the von Neumann architecture of modern computers~\cite{vonN58}.

A modern computer system contains at least five modular components:
the input, output, memory unit, control unit, and computing unit 
(also known as central processing unit). 
In particular, the memory contains stored programs,
which enables the power of modern classical computers to automate the execution of algorithms.
The current paradigm for quantum computing is a quantum-classical hybrid:
the quantum circuit model mainly serves as a quantum central processing unit, 
which is usually classically controlled
with quantum algorithms stored as classical programs. 
There are excessive amounts of classical ingredients which appear inevitable.

There have been persist efforts to go beyond the scope of the circuit model, 
even starting from the beginning of quantum information.
In the setting of quantum Turing machine, 
it was pursued if quantum computing can be fully quantum~\cite{Mye97,Oza98,Shi02},
i.e., with all components including the read-write head, the address of qubits,
programs, control, halt signal, etc being quantum.
It was then discovered that programs cannot be made quantum 
in the sense that once a program is stored as a quantum state, 
it cannot be read out deterministically~\cite{NC97}.
An obstruction for the quantum control over arbitrary quantum operations 
is also revealed lately~\cite{AFC14,TMV18,GST20,VC21}.
These study together with other no-go theorems, e.g., 
Refs.~\cite{Die82,WZ82,BCF96,May97,LC97,BEM98,DP05,ZCC11,CCC+08,EK09}, 
illustrate the sharp distinction between quantum information and classical ones.

The difficulty to formalize a universal quantum computer system is linked to a vital issue of quantum physics.
It concerns if there is a so-called quantum-classical boundary, 
how or where to draw such a boundary~\cite{vonN55}. 
From the modern theory of quantum decoherence~\cite{Zur03},
classicality arises if the coherence of a quantum system is lost or delocalized into another system.
As the elements of quantum information, 
qubits are considered to be the extensions of bits and probabilistic bits (or pbits),
in the sense that a qubit is a superposition of bit values,
and it leads to pbits if it is measured.
This heuristic indicates that, instead of being puzzled by the nature of quantumness, 
the central issue for a proper model of quantum computer system
is to unfold the quantum advantages for various information processing tasks.

In this work, we propose a prototypical model of universal quantum computer system.
The central ingredient is a stored-program scheme based on the quantum channel-state duality~\cite{Cho75,Jam72}. 
The stored quantum programs can be composed together, processed into other ones,
and executed to realize quantum algorithms. 
Our model is not only universal and modular, but also quantum-secure, 
in the sense that it is protected by the uncertainty principle,
which has played vital roles in quantum communication and cryptography~\cite{BB84}.

Our work is made possible based on a few recent progresses.
First, the essence of some no-go theorems becomes clearer,
including the no-programming~\cite{BCA+10,KPG19,YRC20}, no-control~\cite{GST20}, 
and the incompatibility between transversality and 
universality of logical gates~\cite{HNP+17,FNA+19,WA19,WZO+20,KD20,ZLJ20,YMR+20,WWC+21}.
The linearity of quantum operations and fundamental constraints by the 
uncertainty principle are revealed.
Meanwhile, a scheme using Choi states as stored quantum programs is proposed~\cite{W20_choi},
which turns out to be the proper way to bypass the accuracy constraint by the uncertainty principle.
Also, we recently presented a physical understanding of various universal computing models~\cite{W21_model}.
In particular, we pointed out the relation between the uncertainty principle and logical gates,
and the relation between quantum algorithms and quantum combs~\cite{CDP08a,CDP08,CDP09,GW07,Jen11}. 
These progresses, but not limited to, enable the formation of a universal quantum computer system
with less classical ingredients.

\subsection{Overview of our model}\label{sec:oview}


\begin{figure}
    \centering
    \includegraphics[width=.4\textwidth]{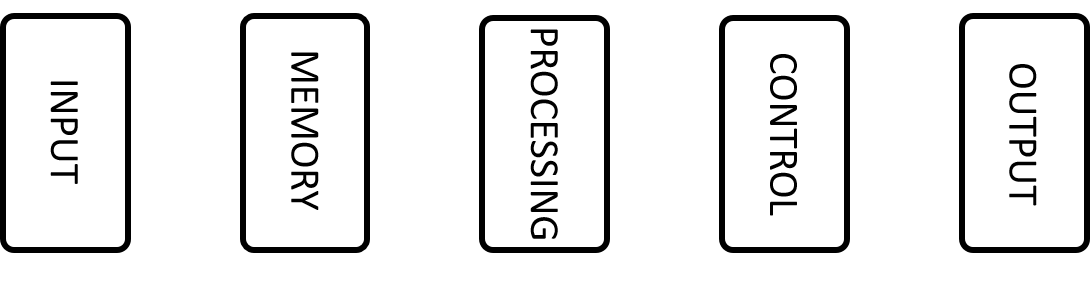}
    \caption{The five primary components of a computer system: 
    the input, output, memory unit, control unit, and central processing unit. 
    Their relations in terms of control and information flows are not shown explicitly here. 
    More details can be found in computer-science textbooks, e.g., Ref.~\cite{NS07}.}
    \label{fig:qcs}
\end{figure}

Here we provide an overview of the model of quantum computer system (QCS),
which shares similarity with the classical case but also shows key distinctions.
As for the classical case, there are five modular components:
the input, output, memory unit, control unit (CU), and central processing unit (CPU). 
See Fig.~\ref{fig:qcs}.
We treat the CU as a separate part from the CPU.
We only focus on the functionality of these components, e.g.,
how they work and relate with each other. 
We do not study devices or hardware in this paper which should be specified for a practical QCS. 
This will be further discussed in Sec.~\ref{sec:conc}.

A modern CPU has a few components such as its own memory and control,
but here we treat it as an arithmetic unit and study the quantum version,
the quantum processing unit (QPU).
A QPU contains at least two parts: the qubits required to perform quantum circuits,
and the devices that are needed to realize quantum gates on the qubits and qubits from the memory.
The quantum memory stores data and programs, both are in terms of Choi states~\cite{Cho75,Jam72}.
Quantum programs are stored as many copies of Choi states of quantum operations. 
Quantum data are stored as many copies of Choi states of preparation circuits for states.
In our model, a computation is carried out on a part of memory 
under the control of a quantum control unit (QCU),
which contains a collection of qubits.
A large program is formed by the composition of small programs from the memory.
The initial input state is ``injected'' to the program by measurements. 
The solution to a given problem is obtained by measurements of observable on the final state.


Compared with classical cases, there are two major differences. 
First, due to the uncertainty principle, quantum data can be made secure,
termed as ``quantum-secure'' in this work,
hence cannot be cloned or estimated efficiently. 
Second, quantum measurements are interactive with random outcomes. 
After a computation, 
the stored programs are consumed but can be restored.
Quantum eavesdropper (Eve) or virus can destroy the programs 
by local measurements without knowing the programs, or, 
if powerful, can do joint measurements on a few copies to estimate a program state,
for which the accuracy is limited by the uncertainty principle.
On the contrary,
classical data can be cloned if not encrypted,
and classical measurements (e.g., read-write operations) are usually deterministic. 
This not only harvests quantum evolution as a computational resource,
but also harvests quantum memory and quantum measurements as resources. 


Using stored quantum programs provides advantages 
especially when this cannot be efficiently done classically,
but also leads to a few challenges. 
First, in order to execute a stored program or algorithm exactly and deterministically,
the input to the algorithm is prepared in a heralded way.
The output of an algorithm is required to be expectation values of observables,
instead of being the final state itself.
It appears that quantum computing occurs in Hilbert spaces,
but eventually, the information carried by quantum states has to be readout by measurements, i.e.,
we have to convert qubits into bits to obtain the solution to a given problem. 
This leads to our refinement of universality to be algorithmic (see Sec.~\ref{sec:ing}).
Also due to the requirement of fault-tolerance,
there will be the potential overhead of quantum error correction on the quantum memory (see Sec.~\ref{sec:fault}).
This also makes a quest to the finding of fully or partially self-correcting qubits~\cite{BLP16}.
Our study proves that a QCS, as the analog of the classical ones, can be established in principle.

This work contains the following parts.
In Sec.~\ref{sec:pre} we survey the primary tools including quantum channels,
quantum combs, quantum error correction, the quantum circuit model and quantum algorithms,
and a few no-go theorems and their relations with the uncertainty principle.
In Sec.~\ref{sec:ing} we explain the main ingredients for our model of QCS,
including the stored quantum programs and data, 
an extension of quantum circuits by stored programs, 
which are defined as tailed quantum circuits,
a definition of algorithmic universality to describe the execution of quantum algorithms,
and a description of the quantum control unit.
We further discuss the features and requirements of the model in Sec.~\ref{sec:feature},
and its relation with some other models.
In Sec.~\ref{sec:fault} we show that QCS can be made fault tolerant,
hence completing the formalization of universal QCS.
Finally, in Sec.~\ref{sec:ext} we briefly discuss a few extensions of our study, 
and conclude with more problems related with QCS in Sec.~\ref{sec:conc}. 

\begin{figure}[t!]
    \centering
    \includegraphics[width=0.3\textwidth]{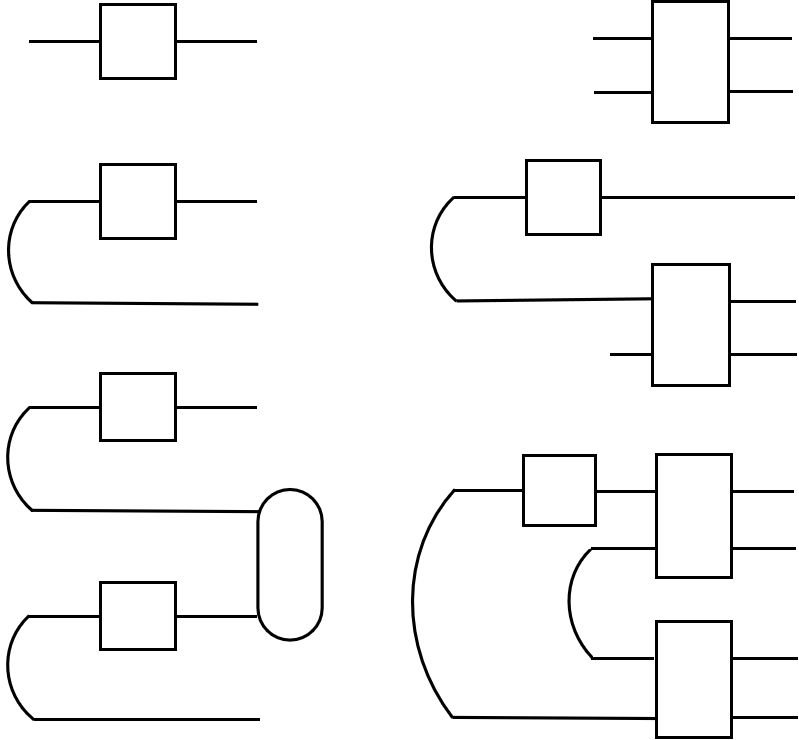}
    \caption{Quantum circuit diagrams (from top to bottom, left to right) 
    for a quantum channel, Choi state, composition of two Choi states, unitary dilation of a channel, 
    initial-state injection scheme on a Choi state,
    and superchannel acting on a Choi state.
    The curved wires are ebits, boxes are quantum operations, their meanings shall be clear from the main text.}
    \label{fig:choi}
\end{figure}

\section{Preliminary}\label{sec:pre}

In this section, we review primary background to make our presentation self-consistent.
Along the way, we clarify facts and draw connections that are relevant to our study of QCS.

\subsection{Quantum operations}\label{subsec:operation}

We consider finite-dimensional Hilbert spaces.
A pure state $\ps$ is an element of a Hilbert space $\C H$ with 
the normalization $\bra \psi \ps=1$ and any global phase is unphysical.
A mixed state, or density operator $\rho \in \C D(\C H)$ is 
a nonnegative semidefinite operator acting on $\C H$ with $\text{tr} \rho=1$,
for $\C D(\C H) \subset \C B(\C H)$ as the convex set of all mixed states,
$\C B(\C H)$ as the space of bounded linear operators acting on $\C H$.
A simple way to distinguish mixed states from pure states is by the purity
\be \text{tr} (\rho^2) \leq 1, \ee
which only equals to 1 for pure states. 
The purity is preserved under unitary evolution,
which is described by unitary operators $U \in \C U(\C H)$ in the unitary group $\C U(\C H)$ acting on $\C H$.
More general evolution is described as completely positive, trace-preserving (CPTP) maps~\cite{Kra83} of the form
\be \C E(\rho)= \sum_i K_i \rho K_i^\dagger, \ee
for $\forall \rho \in \C D(\C H)$, also known as quantum channels, and $K_i$ known as Kraus operators. 
This is also called a Kraus operator-sum representation,
which is not unique due to an isometric degree of freedom.
The minimal number of Kraus operators is the rank of the channel.
A positive operator-valued measure (POVM) is a set $\{F_i\}$ with $F_i \geq 0$, $\sum_i F_i=\I$.
It can be constructed from Kraus operators with $F_i= K_i^\dagger K_i$.
A quantum instrument $\{\Phi_i\}$ is a set of CP maps $\Phi_i$ so that the sum of them $\Phi=\sum_i \Phi_i$ is TP,
and the set of indices $[i]$ is fixed. 
A quantum channel $\C E: \C D(\C H_1)\ra \C D(\C H_2)$ does not need to preserve the dimensions of Hilbert spaces,
but for simplicity, our presentation is made for the dimension-preserving case without loss of generality.
We use $\C E: \C D(\C H)$ to represent a channel $\C E$ that acts on $\C D(\C H)$.

Any quantum channel $\C E: \C D(\C H)$ can be represented as its dual state
\be \omega_{\C E} := \C E \otimes \I (\omega), \ee 
usually known as a Choi state~\cite{Cho75,Jam72}, for $\omega:=|\omega\ket \bra \omega|$,
and $|\omega\ket:= \sum_i |ii\ket /\sqrt{d}$, $d=\text{dim}(\C H)$.
The state $|\omega\ket$ is a Bell state, also known as an ebit.
Kraus operators can be found from the eigenvalue decomposition of $\omega_{\C E}$.
Choi states are bipartite and for clarity, we label them as site A and site B in order.
The partial trace of a Choi state is constrained as $\text{tr}_\text{A} \omega_{\C E}= \I /d$,
and $\text{tr}_\text{B} \omega_{\C E}= \C E(\I) /d$.
Given $\omega_{\C E}$, the action of the channel can be obtained as
\be \C E(\rho)= d \; \text{tr}_\text{B} [\omega_{\C E} (\I \otimes \rho^t) ], \label{eq:readout}\ee
for $\rho^t$ as the transpose of a state $\rho \in \C D(\C H)$.

We see that $\omega$ is the dual state of $\I$.
There is a concise graphical way to illustrate this map, see Fig.~\ref{fig:choi}.
For an operator $A\in \C B(\C H)$, define its vectorization as 
\be |\omega_A\ket:= A\otimes \I |\omega\ket, \ee
which has the property $|\omega_A\ket= \I\otimes A^t |\omega\ket$.
The vectorization or duality is just to bend over the input towards the bottom of the output wire.
This can be generalized if the given system is multi-partite.
For $n$-partite operators, $|\omega\ket^{\otimes n}$ is needed and it holds
\be |\omega_A\ket = A\otimes \I |\omega\ket^{\otimes n}=\I\otimes \tilde{A} |\omega\ket^{\otimes n}, \ee
for $\tilde{A}=R A^t R$, $R$ is the unitary operation that reverses the order of the subsystems.
For the bipartite case, $R$ is the swap operation. 
The relation above can be seen as to shuffle the operator $A$ along the wires from the top to the bottom.
Alternatively, an $n$-partite operator $A\in \C B(\C H)$ can be just treated as single partite, and we can use a high-dimensional Bell state $|\omega\ket \in \C H\otimes \C H$ to define the vectorization.
In this case, $\tilde{A}=A^t$. 
Despite the two choices, we still view the Choi state of an operator as a unique definition.

The general actions on Choi states $\omega_{\C E}\in \C C(\C H\otimes \C H) \subset \C D(\C H\otimes \C H)$
are found to be superchannels~\cite{CDP08a,CDP08,CDP09},
which again can be represented by their dual states.
To describe them, we use the unitary dilation representation of them
which is more appropriate for the quantum circuit model.
For a channel $\C E: \C D(\C H)$, from dilation it can be realized by a unitary $U$ with
\be \C E(\rho)= \text{tr}_a \C U (\rho\otimes |0\ket \bra 0|), \ee
for $\C U$ as the superoperator form of a unitary $U$, 
the trace over an ancilla a at an initial state $|0\ket$,  
which realizes Kraus operators as $K_i=\bra i|U|0\ket$, 
with $\{|i\ket\}$ as states of the ancilla. 
In order to tell channels from superchannels, 
we use a hat on the symbols for superchannels.
A superchannel $\hat{\C S}: \C C(\C H \otimes \C H)$ can be realized as
\be \hat{\C S} (\C E)(\rho)= \text{tr}_a \C V \; \C E\; \C U (\rho\otimes |0\ket \bra 0|),  \ee
for $\rho \in \C D(\C H)$, $\C E: \C D(\C H)$, 
$\C U$ and $\C V$ are unitary,
and a is an ancilla.
Note that the dimension of $V$ can be larger than $U$,
while here we find no need to provide the details of the ancilla.
The Choi state $\omega_{\C E}$ can be made explicit by bending over the
input wires, see Fig.~\ref{fig:choi}, with 
\be \hat{\C S} (\C E)(\rho)=\text{tr}_{\bar{\text{A}}} \C V \otimes \tilde{\C U} 
(\omega_{\C E} \otimes \omega) (\I\otimes \rho^t\otimes |0\ket \bra 0|), \ee
where the support of each operator shall be easy to see hence omitted for simplicity.
The trace is over the subsystems except the top one, A.
The unitary $\tilde{\C U}$ is the transpose of $U$ conjugated by a swap. 
This formalism for superchannels includes channels as a special case with no channel $\C E$ as input and no $\tilde{\C U}$.
We see that in order to change a channel to another one,
we have to use both a pre- and a post- unitary operations, with a memory wire connecting them.
More generally, quantum $n$-combs are defined when $(n-1)$ channels are taken as input sandwiched between unitary operations~\cite{CDP08a,CDP08,CDP09,GW07,Jen11}. 
As such, states, channels, and superchannels are also called 0-combs, 1-combs, and 2-combs, respectively.

In quantum computing, an important type of quantum operations is quantum error correction (QEC),
which corrects errors occurred on quantum error-correction codes (QECC).
A quantum code is often defined by an encoding isometry $V:\C H_L \ra \C H_P$, with $V^\dagger V=\I$,
or by the projector $P=V V^\dagger$ on the code space, $\C C \subset \C H_P$.
A set of error operators $\{E_i\}$ acting on a code $P$ is correctable when
\be P E_i^\dagger E_j P= c_{ij} P, \label{eq:qec}\ee
and $[c_{ij}]:=\rho_P$ can be viewed as a state. 
The correction or recovery scheme, $\C R$, is defined as a set $\{R_k\}$ with $R_k=\frac{1}{\sqrt{d_k}} P F_k^\dagger$
for $d_k$ as eigenvalues of $\rho_P$, and $F_k$ satisfies 
$P F_k^\dagger F_\ell P= d_{k} \delta_{k\ell} P$.
When the error operators form a channel, $\C N$, the logical information is perfectly recovered
$\C R \C N (\ps) =\ps$, $\forall \ps\in \C C$. 
In addition, the condition for error-detection by $P$ is 
$P E_i P= e_i P$, weaker than the QEC condition.

We see that QEC is to find the inverse of a map $\C N$,
which does not exist in general but is possible when its action on a subspace $\C C$ is concerned.
The QEC condition~(\ref{eq:qec}) is actually more powerful:
any set of operators $\{A_i\}$ with each $A_i$ as a linear combination from $\{E_i\}$
can also be corrected by $\C R$. 
This is often known as the linear Kraus-span property as the error operators are considered as Kraus operators for channels.
For codes with tensor-product form $\C H_P=\otimes_n \C H_n$, 
we only need to consider a spanning set of errors for each subsystem $n$,
as others are linear combinations of them. 
For multi-qubit codes, we often consider Pauli bit-flip error $X$ and phase-flip error $Z$ for each qubit,
with $Y$ as a product of them,
and that is sufficient to characterize the primary error-correction features of a code,
such as code distance and threshold.
A QECC is usually denoted as $C=[[n,k,d]]$, which uses $n$ physical qubits to encode $k$
logical qubits, and has a distance $d=2t+1$, namely, 
can correct errors that act on up to $t$ physical qubits at unknown sites. 
There are also approximate QECC which we do not focus on in this paper.


\begin{figure}[t!]
    \centering
    \includegraphics[width=0.3\textwidth]{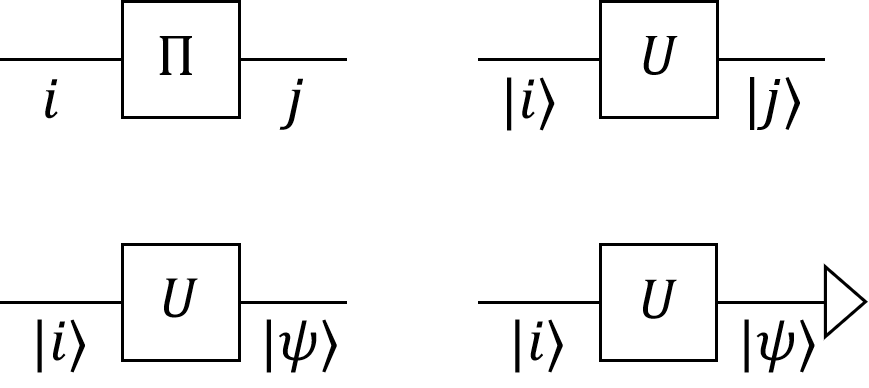}
    \caption{Quantum algorithm and special cases (from left to right, top to bottom): 
    classical one by permutation $\Pi$ from bits to bits (e.g., a bit-string $i$ to $j$), 
quantum one by unitary $U$ from bits $|i\ket$ to bits $|j\ket$ or a nontrivial entangled state $\ps$, 
and with the final bit values read out by measurement. 
This also applies to quantum meta-algorithms, i.e., combs. 
The classical circuit-design algorithm is not shown here, see Fig.~\ref{fig:alg-comb}.}
    \label{fig:alg-ob}
\end{figure}

\subsection{Quantum circuit model}\label{subsec:circuit}

There are a few universal quantum computing models~\cite{W21_model} 
and here we employ the quantum circuit model (QCM) in our study.
In the setting of QCM, universality means that any unitary operator $U\in SU(2^n)$ can be efficiently approximated 
by $\tilde{U}$ to an arbitrary accuracy $\epsilon$. 
The circuit size of $\tilde{U}$ shall be polynomial of $\log \frac{1}{\epsilon}$.

In QCM, a unitary $U$ is realized as a sequence of gates that are available,
and an algorithm is realized by acting $U$ on an input state, followed by a measurement process for readout.
A gate is a unitary operation that can be turned on and off by external control.
A set of gates is called a universal gate set if product of gates from it can approximate any unitary efficiently.
The two well-known examples are the set $\{H,T,\textsc{cx}\}$ and the set $\{H,\it{\textsc{ccx}}\}$
for \be H=\frac{1}{\sqrt{2}}(X+Z),\; T=Z^{\frac{1}{4}},\; \textsc{cx}=P_0\otimes \I+ P_1 \otimes X,\ee 
and \textsc{ccx} as the Toffoli gate $\textsc{ccx}=P_0\otimes \I+ P_1 \otimes \textsc{cx},$ 
$\textsc{cx}$ usually denoted as $\textsc{cnot}$, and $X$, $Z$ as Pauli matrices, $P_{0,1}=\frac{\I \pm Z}{2}$.
The Toffoli gate is known to be universal for classical computation.
We do not need to consider nonunitary gates since, due to the dilation theorem,
any nonunitary quantum channel can be realized by a unitary operator, together with final measurements on ancilla.
When QEC rounds are required,
they can also be realized by unitary operations followed by measurements.

In quantum circuits, quantum gates are causally ordered 
in the sense that the space and time location of a gate is classical
and controlled by a classical computer or system. 
Also the direction of time, namely, the information flow, of a circuit is fixed, 
which is from the given input to the desired output. 
This agrees with the intuition from classical algorithms, 
which can be generalized for the quantum case (see Sec.~\ref{sec:ing}).


\begin{figure}[b!]
    \centering
    \includegraphics[width=0.4\textwidth]{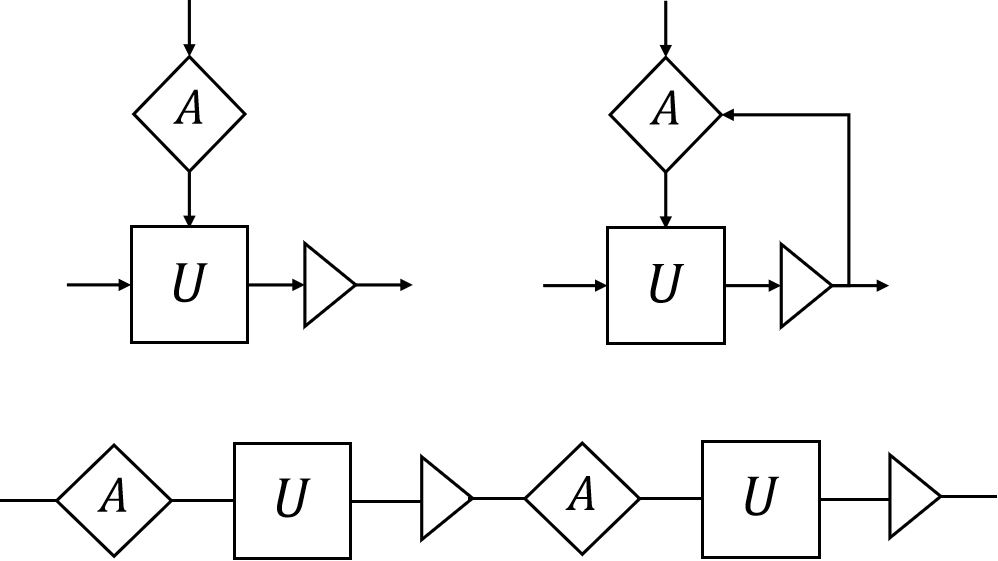}
    \caption{Structures of quantum algorithms. 
    The basic structure (top-left) has a classical algorithm $A$ that designs
    the quantum algorithm $Q$, or labelled by the unitary circuit $U$. 
    It extends to the iterative classical-quantum algorithms (top-right), 
    which can be ``stretched'' into a linear flow (bottom),
    and the most general forms in terms of quantum combs, see Fig.~\ref{fig:comb-comp}.}
    \label{fig:alg-comb}
\end{figure}

Quantum algorithms can be described relative to a universal computing model,
such as adiabatic algorithms and quantum walks,
but they can be translated into the circuit model. 
A quantum algorithm, $Q$, is usually described by a quantum circuit,
together with a proper initial state and readout scheme.
The readout contains the solution $S$ to a given problem, $P$,
which is the input of a classical circuit-design algorithm, $A$.
Namely, $A: P \mapsto [Q]$ for $[Q]$ as a classical description of the quantum algorithm 
$Q: |0\ket\mapsto S$, with $S$ as measurement outcome and $|0\ket$ as an initial state 
required by $Q$. See Fig.~\ref{fig:alg-ob} and Fig.~\ref{fig:alg-comb}.
This framework applies to many quantum algorithms, including quantum phase estimation,
quantum simulation, etc~\cite{NC00}. 
The classical algorithm $A$ is an essential part and often difficult to find,
and it could be limited if $[Q]$ cannot be efficiently described. 



The solution to a problem is encoded in the expectation value
\be o_f:= \text{tr}(\C O \rho_f)\ee
of observable $\C O$ on the final state $\rho_f$.
The measurement scheme of $\C O$ is required to be efficient.
This usually requires running many rounds of the circuit to estimate the observable.
Sometimes the output is just the final state $\rho_f$ without a measurement,
which can be subsequently given into another quantum algorithm,
yet eventually measurement is required to convert quantum states into classical values.
Quantum algorithms are generically probabilistic, namely, 
within a given accuracy $\epsilon$, the approximate solution $o_f$ 
is obtained with a high probability $p$
that can be efficiently boosted towards 1. 
We emphasize here that quantum algorithm is a generalization of probabilistic algorithm,
or in other words, it unifies classical and probabilistic ones.
The output of quantum algorithm is of the form $\text{tr}(\C O \rho_f)$,
which needs both quantum state generation and quantum measurement.

\subsection{Quantum no-go theorems}\label{subsec:no-go}

Although quantum information is an extension of the classical case, 
there are significant differences which are apparently revealed by various so-called no-go theorems. 
The goal of this section is to show that a few fundamental no-go theorems are equivalent,
and the underlying physics is the uncertainty principle.

The no-cloning theorem~\cite{WZ82,Die82} is well known and it states that 
an unknown quantum state cannot be cloned by any quantum operations, 
e.g., from $\ps$ to $\ps\ps$. 
In general, it applies to $n\ra m$ cloning process
\be U\ps^n |\chi\ket=\ps^m, \ee
for $m>n$, $U$ and $|\chi\ket$ are independent from the input $\ps$.
It would violate the linearity of quantum operations.
It was shown to be equivalent to quantum estimation problem~\cite{BEM98}, 
namely,
a perfect quantum cloning machine will estimate an unknown state perfectly,
and a perfect quantum estimation machine will also yield many copies of an unknown state. 
The quantum estimation task can be realized by a unitary operation with
\be U \ps^n |\chi\ket= |\psi'\ket |[\psi]\ket, \ee
for $|[\psi]\ket$ as a bit-string encoding of $\ps$ or the estimated parameters in it.
The states $|[\psi]\ket$ are orthogonal for any pair of input states.
For pure states, the set of states that can be perfectly cloned or estimated
are orthogonal with each other.

The no-programming theorem~\cite{NC97} is also due to orthogonality.
If there is a quantum operation $U$ that realizes
\be U |d\ket |P_G\ket = G|d\ket |P_G'\ket, \ee
for any data state $|d\ket$ and any program state $|P_G\ket$ that 
encodes the unitary operator $G$ serving as a program,
then the set of programmable states $\{|P_G\ket\}$ have to be orthogonal.
Orthogonality implies classicality since an orthogonal set of states can be viewed as a basis of a Hilbert space.
Any superposition of basis states are not programmable.
Note that the output is product states.
The no-programming has also been extended to POVM~\cite{DP05} and the physics is similar.

It is easy to see its relation with the cloning and estimation task.
If $|P_G\ket$ can be cloned or estimated perfectly, 
it would lead to a perfect universal programming machine. 
On the contrary, the state $|P_G'\ket$ is independent of $|d\ket$ 
so it can be recycled~\cite{YRC20} to recover $|P_G\ket$, 
which would lead to the execution of $G$ for arbitrary number of times,
which is a cloning of $G$ or state $G|d\ket$.


Furthermore, it is well understood that quantum estimation can be done probabilistically or approximately,
and probabilistic scheme can also be treated as approximate scheme if the outcomes are mixed together. 
In quantum metrology, it has been established that an arbitrary parameter held by a quantum state or evolution
can only be estimated approximately, and the error is lowered bounded due to the uncertainty principle 
and also the extension of it via quantum Fisher information~\cite{Hel76}.
Although there are many variants of the uncertainty principle based on different operations,
the essential fact it established from the Cauchy–Schwarz inequality
\be |\bra x|y\ket|^2 \leq |\bra x|x\ket|\; |\bra y|y\ket|, \ee
for $|x,y\ket$ from an inner-product space, is that 
the uncertainty of two non-commuting operators on a quantum system 
is lower bounded by the degree of the non-commutativeness of them.
Optimal schemes for universal quantum cloning, estimation, and programming are known~\cite{DBE98,FWJ+14,BCA+10,YRC20},
and the common fact is that, given $n$ copies of the unknown parameter, state, or gate,
the accuracy is no better than the scaling $1/n^2$, 
which is achievable using multipartite entangled states. 
Without entanglement, the accuracy bound reduces to the so-called short-noise limit $1/n$.
The accuracy cannot be exponential (e.g., $2^{-n}$) as that would converge to the perfect case,
hence violating the no-go theorems. 

There is also a no-go theorem in the setting of QECC~\cite{EK09},
which states that the transversal logical gates on a finite-dimensional quantum error-detection code
form a finite group.
We can see the connection with the above no-go theorems from the error-detection condition and output state form.
An encoding operation can be defined by a unitary operator $U: |\ell\ket \ra |\psi_\ell\ket$
that maps logical states $|\ell\ket \in \C H_L$ to the encoded states $|\psi_\ell\ket \in \C H_P$.
For a transversal partition $\C H_P=\otimes_n \C H_n$, 
any transversal unitary operator takes the form 
\be U=\otimes_n U_n. \ee 
Now suppose the states $|\psi_\ell\ket$ are product states $\otimes_n |\psi_n\ket$
instead of being entangled.
If the error-detection condition is required,
then all 1-local states are fixed, which is then impossible to encode logical states $|\ell\ket$. 
If the states $|\psi_\ell\ket$ are product states, 
while the error-detection is only for classical errors,
e.g., for bit-flip errors on any site but not for phase-flip errors,
then we find orthogonal logical states are mapped to orthogonal 1-local states for all sites.
Here we notice the orthogonality and 
this actually leads to classical error-detection codes.
Now if we allow entangled states $|\psi_\ell\ket$,
the error-detection will require the set of transversal logical gates being a finite group.
This is a feature that is not present for other no-go theorems we mentioned above.
If the error-detection condition is dropped, 
then $SU(2^n)$-covariant codes exist~\cite{HNP+17,FNA+19,WA19,WZO+20,KD20,ZLJ20,YMR+20,WWC+21} 
which apparently allow any transversal logical gates.
However, the accuracy of such codes is limited by the uncertainty principle,
and for $n$ transversal parts, the accuracy is upper bounded by $1/n^2$~\cite{KD20,ZLJ20}. 
Indeed, the encoding operation can be viewed as an estimation scheme of $|\ell\ket$,
and the transversal operations are the parallel black-box calls. 
The optimal entangled resource states for estimation are also found to be optimal for $SU(2^n)$-covariant codes~\cite{YMR+20}.

Another notable no-go theorem is the no-control over unknown quantum operations~\cite{AFC14,TMV18},
which forbids the process
\be U_2 (\I\otimes \I \otimes U) U_1= U_3 \otimes CU,\ee
for fixed unitary $U_{1,2,3}$ acting on the tri-partite system,
and any unknown $U$ acting on the third subsystem,
and $CU$ as the controlled-$U$ on the second and third subsystems,
up to an arbitrary global phase on $U$.
It turns out it is of different category from the three no-go theorems above.
It not only violates the linearity of quantum operations,
but also the meaningless of global phases of quantum states or operations.  
The above process would lead to the perfect distinction between a gate $U$ and $-U$,
which is impossible in quantum theory. 
This has been shown as a topological obstruction from the Borsuk-Ulam theorem~\cite{GST20},
and there is no approximate version of the no-control theorem. 
It also forbids the process $|\psi_1\ket|\psi_2\ket \mapsto |\psi_1\ket\oplus |\psi_2\ket$
since it maps global phases of one state to relative phases between the two states. 
In general, this is a ``no-packing'' of unknown quantum operators,
and such packings are not valid operations on Hilbert spaces. 

It turns out there is an easy scheme to overcome the no-control theorem~\cite{AFC14,TMV18,GST20,VC21},
which is to know at least one eigenstate of the gate $U$.
Given the black-box access of $U$, the $CU$ can be realized as
\be CU \otimes \I_a = \textsc{cswap} (\I \otimes U_a) \textsc{cswap}, \ee
for a as an ancilla and $\I_a$ as the completely-mixed state of it, 
$\textsc{cswap}$ as the controlled-$\textsc{swap}$ gate, 
and the input state for the target of $CU$ is the known eigenstate of $U$.
In addition, another case is when the gate $CU$ itself is given with $U$ unknown,
which applies to situations that $U$ has trivial action on a few levels or modes 
that belongs to the whole Hilbert space~\cite{VC21}.
These levels or modes are eigenstates of $CU$ with eigenvalues 1.

\section{Ingredients}\label{sec:ing}

\subsection{Stored quantum programs and data}\label{subsec:sqpdata}

Stored programs and data are important components for the von Neumann architecture of computers.
Despite the no-programming theorem, 
here we present an efficient scheme of stored quantum programs and data, based on the recent study~\cite{W20_choi},
and this is the starting point for our model of QCS.
The quantum memory unit in QCS contains many programs and data, 
each of which prepared many copies and stored with definite addresses in the memory unit.
The query of quantum memory and computation with them is efficiently controlled by the quantum control unit.
For simplicity, we focus on unitary evolution of pure states, 
which can be quite straightforwardly extended to non-unitary evolution of mixed states.

For a pure state $\ps\in \C H$, a simple scheme is just to prepare $\ps$ and store it as quantum data.
While here we introduce a different scheme that will be employed in our model.
We define its computational preparation circuit as $U$ with $\ps=U|0\ket$, 
for $|0\ket$ as a computational basis state of $\C H$. 
The $U$ is not unique but can always be chosen properly.
We define quantum data of $\ps$ as the Choi state $|\omega_U\ket$ of its preparation circuit $U$. 
For a $n$-qubit state, it needs $2n$ qubits to store it as quantum data.
In general, this storage can be made arbitrarily accurate if the accuracy of $|\omega\ket$ and $U$ can be guaranteed. 
On the contrary, we can store a state $\ps$ directly as classical data, denoted as $[\psi]$.
For generic $n$-qubit states, this is to convert all the amplitudes $\psi_i=\langle i\ps$ into bits,
which cannot be efficient with respect to $n$. 
Only special types of states such as stabilizer states~\cite{Got98} 
can be described efficiently using bit strings on classical computers.

A quantum program or algorithm is defined by a unitary operator,
while often also requires special initial states as input and special measurements as readout scheme. 
For simplicity, we treat the input as a part of data, and readout as a separate stage.
Namely, a quantum program merely refers to a unitary operator $U$.
Similar with quantum data, a stored quantum program $|\omega_U\ket$ is the Choi state of $U$.

From the relation~(\ref{eq:readout}), 
the action $U\ps$ can be realized by the measurement of $\ps$ on site B of $|\omega_U\ket$~\cite{W20_choi}. 
Define a binary projective measurement $\{P_0,P_1\}$ with 
\be P_0^t= \ps\langle \psi|, \; P_1=\I-P_0,\ee
for $P_0^t$ as the transpose of $P_0$, and the measurement outcomes 0 and 1 are recorded.
When the outcome is 0, the state on site A is $U\ps$.
When the outcome is 1, the state on site A is $\I - U\ps\langle \psi| U^\dagger$.
This is enough for the readout of computational result $\text{tr}(\C O\rho_f)$,
which is encoded as an observable $\C O$ on site A, for $\rho_f=U\ps\langle \psi| U^\dagger$
and the value $\text{tr}(\C O)$ is efficiently computable required by the readout scheme 
(see Sec.~\ref{subsec:univer} for more discussions). 

When the input $\ps$ is stored as the quantum data of its preparation circuit,
we need the composition of Choi states. 
Given two Choi states $|\omega_{U_1}\ket$ and $|\omega_{U_2}\ket$, 
the symmetry-based quantum gate teleportation~\cite{W20_choi} leads to the state $|\omega_{U_2U_1}\ket$
or $|\omega_{U_2^tU_1}\ket$ in a heralded way. 
Namely, a qubit ancilla is used to encode the outcomes of Bell measurement being trivial (with no byproduct) and non-trivial 
(with Pauli byproducts), recorded by 0 and 1.
The case of being 0 requires no further action but only teleports $U_2^t$, 
while the case of being 1 can teleport $U_2$ but need a correction rotation 
$U_{2;\text{adj}}$ as the adjoint representation of $U_2$.
As a unitary matrix can be decomposed as a product of two symmetric unitary matrices,
a program can be stored by two Choi states,
and the composition 
\be U_\text{UQT} |\omega_{U_1}\ket |\omega_{U_2}\ket= |\omega_{U_2 U_1}\ket\ee
will avoid the transpose (of $U_2$) and becomes deterministic. 
In this work, we name the symmetry-based quantum gate teleportation as a $SU(d)$-covariant or universal quantum teleportation (UQT) since it can teleport any unitary gate, which eventually guarantees the universality of our stored-program scheme.
The UQT serves as the composition operations and is denoted by a rounded box in the circuit, see Fig.~\ref{fig:choi}
(or by distinguished wires in Fig.~\ref{fig:tail2} or circles in Fig.~\ref{fig:highd-comp}).

We see that after the execution,
the quantum program is destroyed, i.e., overwritten by trivial bits. 
Even with many copies of a quantum program, 
it can only be used for a finite amount of runs.
This seems to be a drawback of quantum data, 
but it turns out to be the opposite.
Quantum programs are secure and expensive,
which stand as key distinctions from classical programs and memory,
which can be perfectly cloned and reused forever, 
if no encryption or subscription required (also see Sec.~\ref{sec:feature}).
After a quantum program is executed and overwritten by trivial bits,
it shall be restored by downloading it.

\begin{figure}[t!]
    \centering
    \includegraphics[width=0.4\textwidth]{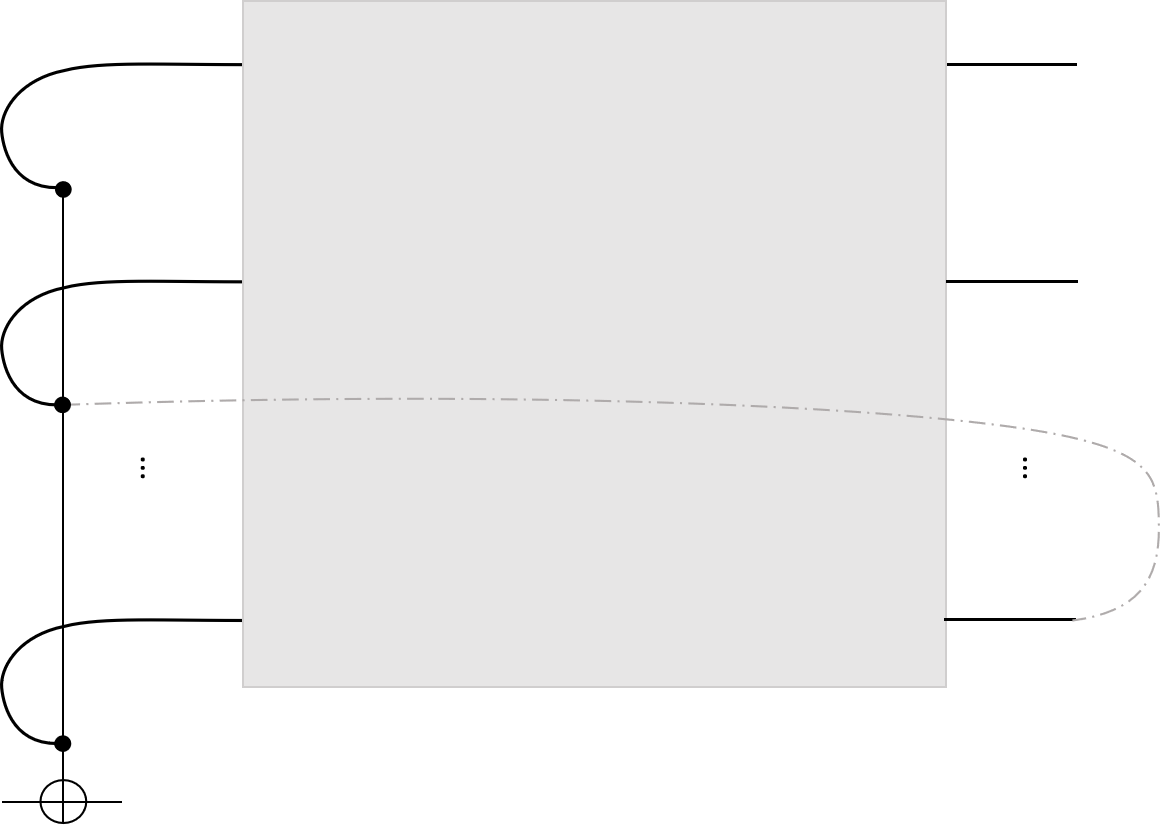}
    \caption{Schematics of a tailed quantum circuit. 
    The shaded box is a unitary operator. 
    The curved wires contain the tails, which are arranged at the left side for the input.
    The multiple-controlled NOT gate (i.e., a $n$-fold Toffoli gate) is for the initial-state injection.
    The dotted gray wire is an example of a contraction on a head-tail pair.}
    \label{fig:tail}
\end{figure}

A program shall not only be stored in memory,
but also shall be downloadable from the internet.
Although the internet is an extra object of computer system,
here we present a scheme to download quantum programs from an anticipated quantum internet~\cite{WEH18}.
The downloading operation for classical case contains the cloning of classical data
and overwriting some amount of bits in the memory.
For the quantum case, this cannot be done since quantum data cannot be cloned. 
Quantum programs are provided by quantum software vendors,
who wish to keep the programs unknown to the agent.
The quantum internet requires the communication of qubits.  
However, the Holevo's bound asserts that a qubit can communicate at most a single bit~\cite{Holevo82},
consistent with the uncertainty principle and the no-cloning theorem.
For a quantum program $U$,
flying qubits such as photons can be prepared as the entangled state $|\omega_U\ket$ itself in principle,
but the agent cannot use them to recover the quantum program $U$ efficiently in its quantum memory.

Fortunately, quantum cryptography~\cite{BB84} shows that bits can be securely communicated using qubits.
Meanwhile, a quantum program $U$ can be described by its classical information efficiently, 
if it is given as a sequence of gates composing a quantum circuit, $\prod_\ell U_\ell$.
Namely, given a universal gate set, the type of each gate (e.g., $H$, $T$, $\textsc{cx}$) can be encoded by two bits.
The space-time location of an elementary gate can be encoded by bits efficiently.
Therefore, the classical information of $U$ contains the bits for the types and space-time locations of the gates.
For a circuit which is efficient with respect to the number $n$ of qubits and accuracy $\epsilon$,
the bit-string description of the gate sequence of $U$ is efficient.
Note that although it can be used to construct $U$, yet as a single matrix, 
$U$ cannot be stored efficiently by bits in general. 
The bit-string description of $U$ can be encrypted by a quantum software vendor,
and securely communicated using qubits (or even using the post-quantum encryption~\cite{BL17}).
After receiving the bit-string description, the agent, which is a quantum computer and
protected against by the encryption, can apply the gate sequence to restore the quantum program.
The quantum program remains unknown to the agent,
hence the program is securely downloaded and the quantum program is secure. 
In all, we demonstrated the viability of stored quantum programs,
and we also remark that practical security is far more complicated 
than the primary scheme we presented here.

\subsection{Tailed quantum circuits}\label{subsec:tqc}

\begin{figure}[t!]
    \centering
    \includegraphics[width=0.35\textwidth]{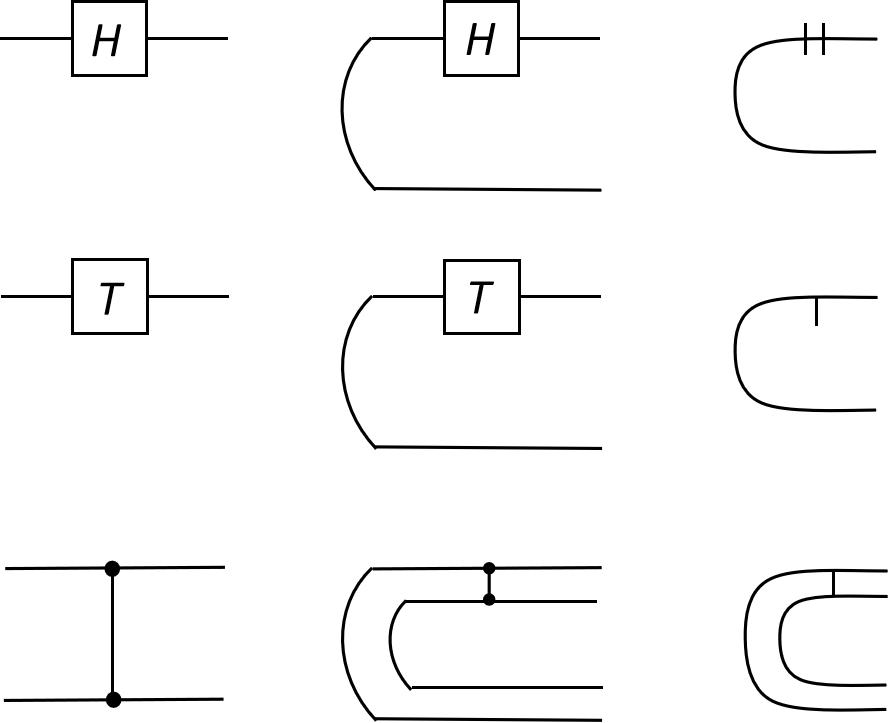}
    \caption{The elementary quantum gates $H$, $T$, and $\textsc{cz}$ (left), their Choi states (middle), and simplified notations (right).}
    \label{fig:tail1}
\end{figure}

\begin{figure}[b!]
    \centering
    \includegraphics[width=0.4\textwidth]{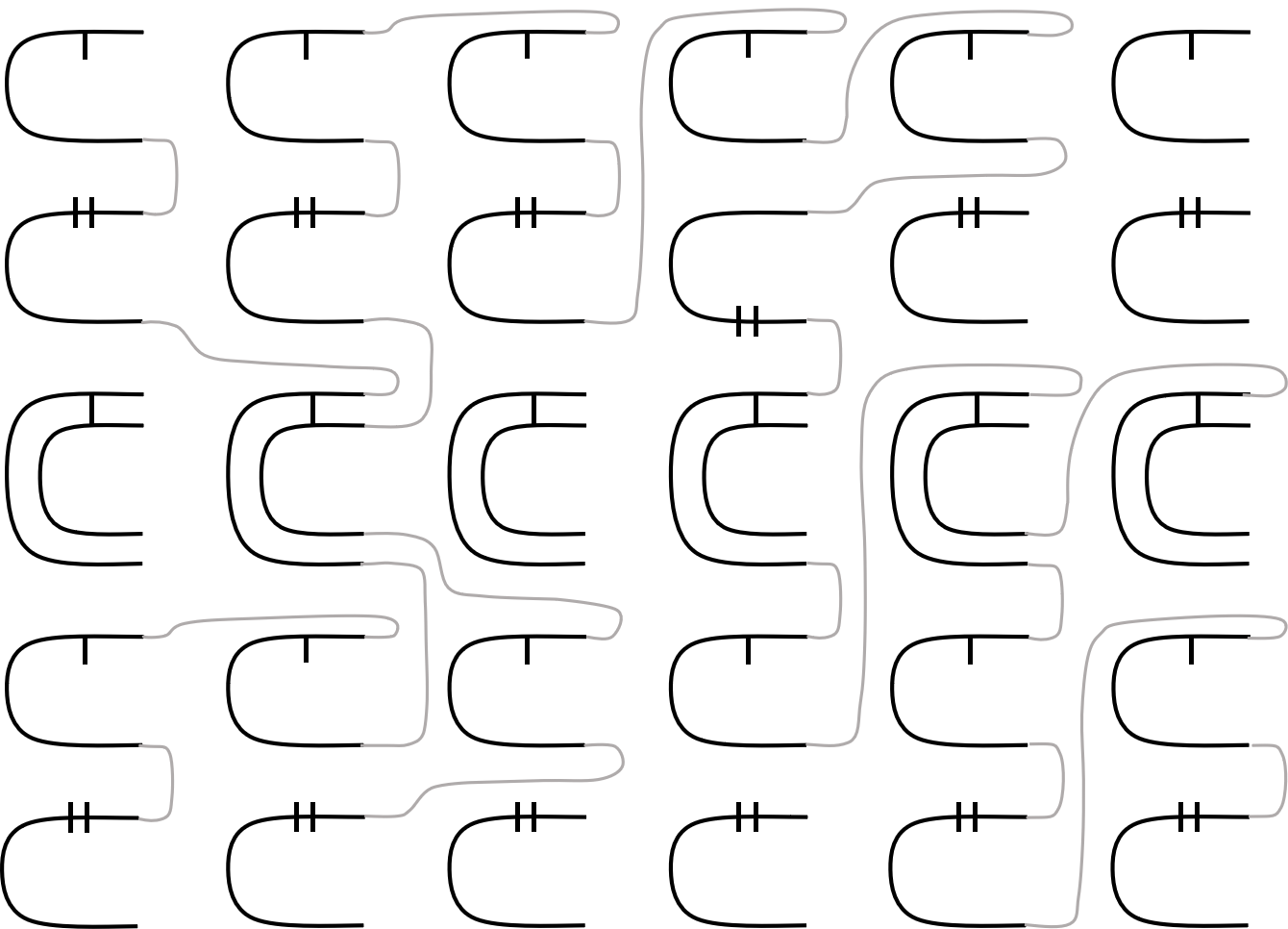}
    \caption{An example of a quantum circuit realized by composition of elementary quantum programs for $H$, $T$, 
    and $\textsc{cz}$ stored in the quantum memory. The compositions are denoted by gray wires.}
    \label{fig:tail2}
\end{figure}

In accord with the stored quantum programs,
we need to introduce a modification of the standard quantum circuit model, 
termed as tailed quantum circuit model.
Recall that a quantum program $|\omega_U\ket$ is a bipartite state,
and we name the subsystem acted upon by $U$ as `head', and the other as `tail'.
A tailed quantum circuit is a quantum circuit $U$ of a sequence of quantum gates
that act on a few qubits and the heads of a few ebits,
while each tail is left unchanged. 
Without qubit input,
a tailed quantum circuit is just a program state $|\omega_U\ket$.
The input for a program is injected to the circuit by making measurements on a few tails. 
The readout is specified by quantum measurement and supported on a few heads.
The input is always carried by tails, output carried by heads.
See Fig.~\ref{fig:tail} for an example.


Due to ebits, we can apply some novel operations with tailed quantum circuits.
Given a collection of elementary tailed quantum circuits,
they can be connected in series and parallel to form larger programs.
The elementary tailed quantum circuits are those for elementary gates from universal gate sets.
Familiar gates are Pauli gates $X$, $Y$, $Z$, and $H$, $T$, $\textsc{cz}$, $\textsc{cx}$, $\textsc{ccx}$, $\textsc{ccz}$, etc.
We focus on $H$, $T$, and $\textsc{cz}$ as examples, see Figs.~\ref{fig:tail1} and~\ref{fig:tail2} 
for an example with simplified notations. 
They are all symmetric matrices, so they can act on either the upper or the bottom wire,
but as a convention, we choose the upper one as the head.
Using the UQT defined in subsection~\ref{subsec:sqpdata},
smaller tailed circuits can be composed into larger ones, 
just as the composition of gates in a usual quantum circuit.

A large program $U\in SU(d)$ can be stored with a `bold' high-dimensional tail of dimension $d$,
or with a qubit tail for each qubit in the system when $d$ is converted to $2^n$ for a number $n$.
When there are multiple tails and heads, contraction (or fusion) of a pair can be made 
by a Bell measurement on them. 
A pair can be of any form, head-head, tail-tail, or head-tail.
Cares are needed to properly choose the time flow 
after the input and output measurements are being made
to avoid backward flow in time or closed time loops.
The application of the contraction will be further discussed in section~\ref{sec:ext}.

As quantum measurement outcome is random, 
a tailed quantum circuit can be employed to sample a collection of circuits.
For each ebit, a measurement of Pauli $Z$ on its tail injects either state $|0\ket$ or $|1\ket$ with equal probability,
or in other words, a Pauli $X$ byproduct that cannot be corrected.
For the contraction of a pair of head and tail, the Bell measurement yields a connected wire but with Pauli byproduct $X$, $Y$, or $Z$
that cannot be corrected, neither.
The measurement outcomes are recorded so can be used to sample circuits with different initial states 
and final measurements.
However, this also means that an initial state, e.g. a computational state $|\vec{0}\ket=|00\cdots 0\ket$,
cannot be prepared deterministically. 
This is indeed true since a qubit state cannot be obtained deterministically from an ebit.
To execute an algorithm on a tailed quantum circuit,
we need a different input-injection scheme that is studied in the next subsection.


\subsection{Algorithmic universality}\label{subsec:univer}

\begin{figure}[b!]
    \centering
    \includegraphics[width=0.28\textwidth]{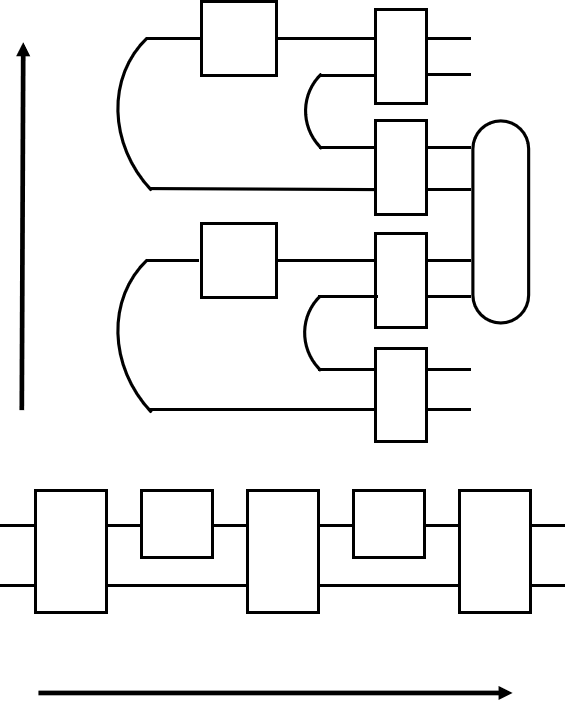}
    \caption{A quantum comb (bottom) and its realization by the composition on Choi states (top). 
    The information flows towards right for the circuit of a comb, while flows upwards for the composition.}
    \label{fig:comb-comp}
\end{figure}

Here we define a so-called algorithmic universality relative to tailed quantum circuits, 
and show how to run quantum algorithms with tailed quantum circuits.
Basically, it needs initialization, unitary evolution, and the readout measurement.
The unitary operation is given as the program state.
Suppose the input is required to be $|\vec{1}\ket$,
which is easier to illustrate than $|\vec{0}\ket$.
We define a binary measurement $P=\{P_0,P_1\}$ with $P_0=\I-P_1$ and $P_1=|\vec{1}\ket\bra \vec{1}|$.
A qubit ancilla initialized at $|0\ket$ is needed to realize it.
For $n$ qubit tails, a $n$-fold Toffoli gate~\cite{BBC+95} is needed to copy the AND of all qubit values to the ancilla (see Fig.~\ref{fig:tail}).
The $Z$ measurement on the ancilla realizes $P$: 0 for $P_0$, 1 for $P_1$.
The $n$-fold Toffoli gate can be decomposed into a product of polynomial number of elementary gates.
In particular, it can be decomposed as a cascade of $n-1$ Toffoli gates with $n-1$ qubit ancilla.
Furthermore, it is easy to see the probability for $P_1$ is $2^{-n}$, which is tiny.
In other words, most of the time $P_0$ is realized.
In this case, the output is $o_f':=\text{tr} (\C O)- \bra \psi_f | \C O |\psi_f\ket$
for $o_f=\bra \psi_f | \C O |\psi_f\ket$ as the true desired output, with $|\psi_f\ket=U|\vec{1}\ket$.
The value of $o_f'$ can be estimated by running the algorithm multiple times,
as usually to be the case for quantum algorithms.
Given that $\text{tr} (\C O)$ is easy to compute,
then $o_f=\text{tr} (\C O)-o_f'$ is obtained with high probability.

The above demonstrates that any quantum algorithm that can be realized on a usual quantum circuit
can also be realized on a tailed quantum circuit.
This proves the universality of the tailed quantum circuit model,
which we term here as an \emph{algorithmic universality}, 
since it is defined in the setting of quantum algorithms.
In passing, probably a better term could be observational universality since it is due to the ability to compute observable values,
or a weak universality based on the weak operator topology on a Hilbert space~\cite{W15_QS}.
Furthermore, the algorithmic universality is actually more complete than the usual notion of universality,
which does not take account of the cost of readout explicitly.
For instance, some readout scheme cannot be done efficiently (with respect to $\epsilon$), 
such as the estimation of unknown gates~\cite{BCA+10} and an approximate stored-program scheme~\cite{YRC20},
which actually reduce the universality to a quasi universality~\cite{WZO+20}.
The algorithmic universality guarantees the universality of a model to realize quantum algorithms.




The study above also extends to more general quantum algorithms.
Just as quantum combs are able to describe general quantum operations,
it is natural to see that they also describe more general types of quantum algorithms~\cite{W21_model}. 
A quantum comb takes a set of quantum objects $\{Q_n\}$ as input,
but uses quantum operations to change them into a desired output,
with a quantum adversary as resource.
This forms a quantum meta-algorithm that designs a quantum algorithm by another quantum algorithm (the comb).
The input $\{Q_n\}$ can be given as unitary oracles or known as black boxes. 
If they are given as stored programs $\{|\omega_{Q_n}\ket\}$,
we can use the composition scheme to connect them and form a comb.
That is to say, we can implement a quantum comb using a stored-program scheme,
see Fig.~\ref{fig:comb-comp}.
Each unitary operator in the comb can be decomposed as a product of two,
then an input object $Q_n$ surrounded by two unitary operators is a block for a superchannel $\hat{\C S}_n$.
Now for each $|\omega_{Q_n}\ket$ we first apply a superchannel $\hat{\C S}_n$ on it,
and then we apply the composition of them in sequence to form the comb. 
In addition, the classical-quantum hybrid algorithms (see Fig.~\ref{fig:alg-comb}),
which feed measurement results from $Q$ to $A$,
can be viewed as a special case of classical combs with the sequence of $Q$s as its input 
and the measurements and $A$ as the comb.

\subsection{Quantum control unit}\label{subsec:control}

\begin{figure}[t!]
    \centering
    \includegraphics[width=0.4\textwidth]{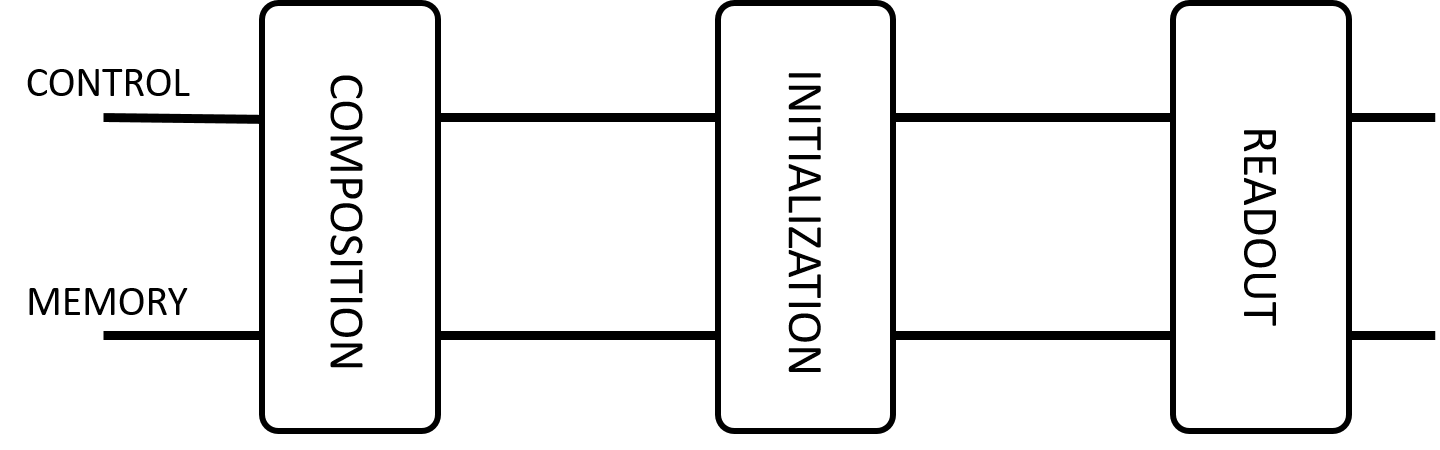}
    \caption{The schematics to show the role of quantum control in a computation.}
    \label{fig:cont}
\end{figure}

The control unit (CU) is an important component of a computer system.
In general, it interacts with all other components of a computer system,
but usually it does not carry final solution to the given problems.
It controls or monitors the procedure and progress of information processing during computation.
For instance, it needs to guarantee that instructions from any algorithm are executed in the right order.
Also it needs to guarantee that data and programs are read and written at the right place in the memory.
Besides universality, there is also a feature of modularity such that CU and CPU are independent components of a computer system. 
Here, we lay out the basic principle to design a quantum control unit (QCU), 
without going into details of the so-called control bus and data bus~\cite{NS07}.
We find there are freedoms that can be explored for its design. 

We first analyze the idea of control and quantum control from the viewpoint of algorithms.
A control scheme is a procedure that aims to achieve a goal by applying a set of control items on a system. 
It can be understood algorithmically,
but with different goals and costs from algorithms.
Quantum control often uses a set of quantum operators to control a scheme to achieve an operational goal~\cite{BCR10}.
The control operators may be supported by the target system itself, or not.
It is straightforward to see that 
a general quantum control can be described as a quantum comb,
with the adversary as the control.
The controller becomes entangled with the target system in general,
but it shall not contain final solutions.
For instance, a common task in quantum control is to use external fields to control the states of atoms or electrons,
which is often semi-classical since there is no entanglement between the fields and the systems.
There are also controls with feedback. 
The simplest example is the $\textsc{cnot}$ gate, which is an entangling gate.
In terms of Pauli operators, the target Pauli is also copied to the control qubit.
Another example is using measurement on the target, 
and the outcome is feed back to the control, and this basically behaves as an iterative algorithm.

In our model, the QCU are qubits and their interactions with other components.
This includes the control of
the composition of programs, the initial-state injection measurement,
and readout measurement, see Fig.~\ref{fig:cont}.
We find there is a freedom to choose for the nature of the control.
The minimal type is classical, namely, 
any quantum algorithm can be monitored classically.
The classical control does not carry quantum information.
On the other hand, the control can be as quantum as possible,
and this is described as quantum combs.
As a result, the control scheme could be an inevitable part of a quantum algorithm, 
and the design of quantum algorithms would involve a significant part for the controller.
Actually, some quantum algorithms can be understood in this way,
such as the quantum switch~\cite{CAPV13} and 
the linear combination of unitary operations~\cite{Long11,CW12}.

Furthermore, whenever the controller participates the computation in a nontrivial way,
the corresponding comb can be stored as a set of programs again. 
Then another level of control is required to realize this comb by composition.
This eventually reduces the control to classical ones. 
Such a reduction of the control sequences or levels is similar with phenomena in other topics, 
such as algorithms and quantum measurements.
For algorithms, we can always add a pre-algorithm that designs the algorithm,
no matter it is classical or quantum.
For quantum measurement, there requires a cut to tell how a special value of an observable is obtained
from a set of possible values, as originally studied by von Neumman~\cite{vonN55}.

Therefore, we require a minimal or basic set of functions of a QCU.
We shall make the QCU modular so it can be used for all quantum targets easily. 
The basic operation is the quantum control of quantum operation,
which is harder than classical control of quantum operation,
but easier than general entangling quantum operations. 
As the study of the no-control over unknown quantum operations in Sec.~\ref{subsec:no-go} reveals that,
an unknown quantum operation as a grey box can be controlled by qubits.
Using qubits as control is a resource that can be explored,
just as the quantum algorithms mentioned above demonstrated.
With bits, each run of a quantum algorithm is a unitary $U$ applying on a fixed initial state $|\psi_i\ket$
and a fixed readout $o_f$. 
With qubits, there could be superposition of quantum algorithms by applying different composition and measurements
conditioned on the control qubits, i.e., 
different unitary, initial states, and readout can be realized in parallel and interfere.


\section{Features}\label{sec:feature}

Our model of QCS requires both qubits and ebits (i.e., Bell states), 
Bell measurement and its generalization,
and both unitary evolution and measurement play vital roles.
The input does not have to be fixed at the beginning. 
The readout scheme avoids the problem in the setting of no-programming theorem,
which requires input as a separate state and the output is a state instead of measurement outcomes.
A nontrivial POVM readout can be simulated by a unitary and a simple projective measurement from dilation theorem.
The quantum control in general requires multiple-controlled gates,
which can be decomposed into elementary gates but that complicates the control process.
The usage of ebits as quantum memory signifies the distinction 
between quantum information and classical information. 
It is protected by the uncertainty principle. 
Below we discuss our model in more details to
reveal its relations with other computing models or schemes,
and its overall features, requirements, and limitations.



First, we clarify the notion of a computer system.
The universal computing model mainly refers to the framework to process information,
while a computer system mainly refers to the architecture to 
divide and combine various aspects of information processing.
A quantum computer system is a particular quantum system that is designed for computing tasks.
Together with our formulation of universality,
we emphasize the difference between quantum computing 
and a usual quantum evolution that occurs in nature everyday.
Quantum computing is usually formulated just as instrumentalists do:
prepare an initial state, let it evolve, and then observe via measurement.
This is an analog description. 
Instead, our study shows that quantum computing is more and needs to be almost fully digital.
The states need to be encoded as qubits, programs encoded as ebits,
and the final solution $o_f$ is carried by an observable, 
which can in principle be converted to bit strings 
such as using quantum amplitude estimation algorithm~\cite{BHM02}.


As has been mentioned, 
the current QCS is a quantum-classical hybrid, or known as quantum random-access machine~\cite{Knill96},
which contains both classical and quantum registers. 
The quantum data on the quantum registers would not be entangled with the control or program.
In the original sense~\cite{Mye97,Oza98,Shi02}, our model of QCS is also not fully quantum, 
despite the usage of quantum control and quantum programs.
For instance, we do not need a halt qubit to signal the end of a computation.
Instead, a computation or an algorithm ends with measurements. 
Actually, the vital thing is not whether it is fully quantum or not;
instead it shall be its computational power and physical flexibility, 
such as being local and modular.
As we have discussed for the quantum control,
we can always shift the boundary between the quantum and classical parts.
Also we currently do not require the ability of the quantum query of all quantum data in superposition,
as that in the scheme~\cite{GLM08} for a quantum random-access memory. 


The QPU in our model is described by the quantum circuit model,
which can be replaced by other universal models,
such as quantum Turing machine and quantum cellular automata,
which are the quantum versions of their classical analogs~\cite{W21_model}.
Meanwhile, if we view the composition as a whole, 
it prepares matrix-product states or tensor-network states, 
which are universal forms of quantum states~\cite{AKLT87,PVW+07,Sch11}. 
This is linked with a local model of quantum Turing machine~\cite{W20_Turing} 
which prepares matrix-product states assisted by a quantum adversary.
But here it is the quantum adversary (or edge) that carries the logical information.


Our model can be viewed as an extension of quantum communication and cryptography
by the universal computation on ebits and Choi states, besides qubits.
The computation with Choi states introduces input by measurements,
and mainly concerns the final expectation value $o_f$ of observable as output.
If the input is treated as an initial state and carried by a separate system from the program,
one has to use a highly entangled optimal program state, e.g.,
a generalized Choi states and global covariant measurement for the retrieval operation~\cite{BCA+10,YRC20,DBE98},
which is constrained by the uncertainty principle. 
That is to say, our stored programs are protected by the uncertainty principle, namely,
if an Eve or virus tries to obtain a stored program,
a global measurement on many copies of it has to be applied with limited accuracy.




Our model can also be seen as an extension of the measurement-based quantum computing (MBQC)~\cite{RB01}. 
In MBQC, the resource state does not contain the program;
instead the program is the measurement bases. 
Each measurement is one-local and it is based on the $U(1)$ symmetry of teleportation~\cite{Wang19b}. 
The measurements are adaptive in order to avoid the byproduct from teleportation.
The original universal blind quantum computing~\cite{BFK09} is based on MBQC, 
which achieves blindness or security via pbits to hide measurement bases.
The security relies on the computational difficulty of searching for the right program over a large set of possible ones.
In a fusion-based quantum computing~\cite{BBB21}, Bell measurements are used to grow graph states or stabilizer states,
but there is no stored quantum programs.
In our setting, we use two-local covariant measurements as an extension of Bell measurements,
which do not contain the programs;
instead, the programs are ``pre-stored'' as quantum states.
Our model has a close connection with valence-bond solids (VBS)~\cite{AKLT87} (see Sec.~\ref{sec:ext} for more details).
As a result, we establish our model of QCS to be 
digital, universal, modular, and quantum-secure,
and can also be made fault tolerant.

\section{Fault tolerance}\label{sec:fault}


In this section, we show how the QCS is consistent with the requirement of fault tolerance.
The quantum fault-tolerance or threshold theorem states that 
universal quantum computation on logical qubits can be realized
if the physical error rate for each logical qubit is below a threshold 
that is determined by the QEC~\cite{NC00}.
The QEC itself could also be noisy, whose effect is to reduce the threshold value.
Physical gates to realize logical ones may also be imperfect, 
which are treated as perfect ones followed by noises.
The fault tolerance is implicitly required by universality since
between any two logical gates, $U$ and $V$, 
QEC is needed to ensure the identity gate $\I$ to form $U\I V$.
For a QCS, it also needs to replace qubits by logical qubits
for all its components, including the memory, control, gates, measurements, etc.
We find that the main issue is the fault tolerance of the composition operations.

For the quantum memory, an ebit is replaced by a logical ebit,
which can be efficiently prepared.
If a code is defined by an isometry $V$, the logical ebit can be obtained as 
\be |\omega\ket_L=V\otimes V|\omega\ket, \ee
for $|\omega\ket$ as the encoded ebit. 
A logical stored program state $|\omega_U\ket_L$ is obtained by applying a logical gate $U$ on $|\omega\ket_L$.
For a code $[[n,k,d]]$, logical gates $U$ commute with its projector $P$, $[U,P]=0$.
For different codes, the form of $U$ varies significantly.
The composition of two logical program states $|\omega_U\ket_L$ and $|\omega_V\ket_L$ is done by 
the logical version of the composition scheme,
which is a UQT on $2n$ physical qubits.
A qubit ancilla is needed for a composition,
which in principle can use a different code.

In the composition scheme, logical gates need to be symmetric to avoid transpose of gates.
This is a nontrivial requirement on logical gates.
We find a concise scheme that satisfies the requirement.
First notice that tensor product of symmetric gates are still symmetric. 
So it is easy to teleport transversal logical gates that are symmetric.
It turns out the elementary gates are all symmetric
such as logical $H$, $T$, and $\textsc{cz}$.
Then we can use the scheme of code switching~\cite{PR13} to combine 
the transversal logical gates from different codes.
For instance, if there are two codes $C_1$ and $C_2$ that can be fault-tolerantly
switched into each other by measuring their stabilizers,
then their transversal logical gates can be combined together,
even achieving universality~\cite{Bom15}.
Then we can apply code switching and composition on them to form large programs.

For a composition, a gate $U_\text{adj}\in SO(N^2-1)$ for $N=2^n$ is needed to teleport a logical gate $U$.
Such gates $U_\text{adj}$ can be efficiently done.
However, they are not logical gates in general for the given codes.
Actually, a sequence of compositions can be viewed as a code concatenation procedure,
with the given codes as the inner codes,
and the outer code is an $SU(N)$ VBS edge code~\cite{WZO+20}.
Now the fault-tolerance of the composition is the fault-tolerance of the concatenation procedure,
and this is also a common issue when just preparing a code.
For a code defined by an encoding isometry $V$, 
there could be noises during the encoding itself. 
It could be hard to do QEC before the code is prepared,
and the threshold theorem implicitly assumes that this can be done.
For codes that the code projector $P$ is composed of a product of smaller projectors $P=\prod_n P_n$,
e.g., for stabilizer codes~\cite{Got98} and codes defined by frustration-free Hamiltonians,
each projector $P_n$ is prepared gradually, hence can be used for QEC during the encoding or concatenation.

Despite the similarity,
there is also an important difference between the composition and code concatenation.
In order to run a quantum algorithm,
the VBS outer code does not have to be prepared on the first hand.
Instead, we can do initial-state injection first, 
and then composite the first logical gate, then the second logical gate, and so on.
The VBS outer code only appears in the time direction.
Actually, this is similar with MBQC~\cite{RB01} and it reduces the idler time and QEC cost.
To ensure fault tolerance,
QEC is performed on the inner codes before and after the teleportation of each logical gate.
Therefore, the fault tolerance is mainly determined by the inner codes for logical qubits and ebits.

\begin{figure}[t!]
    \centering
    \includegraphics[width=0.4\textwidth]{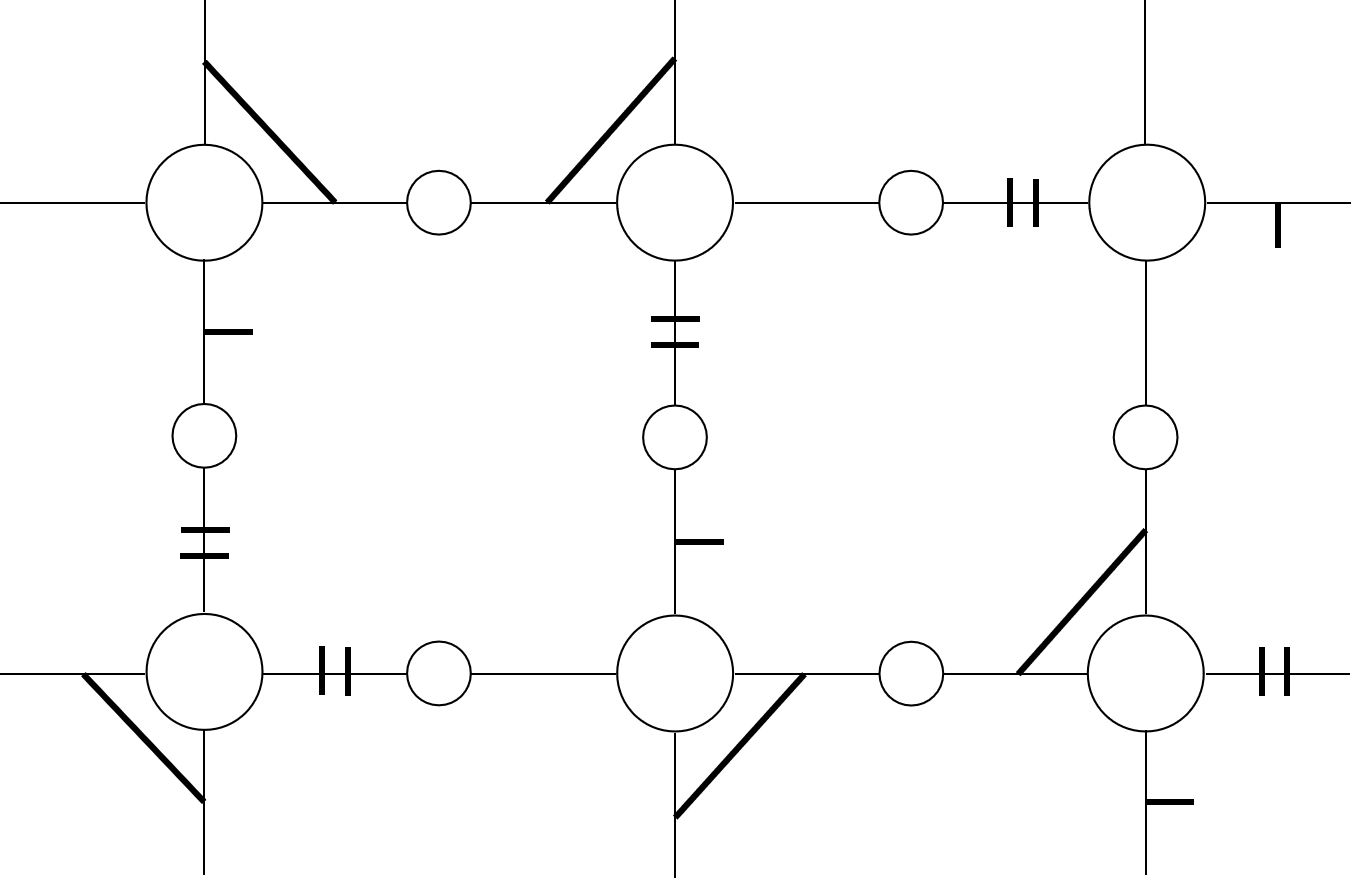}
    \caption{An example of a portion of a high-dimensional composition of quantum programs.
    The big circles are $SU(4)$ compositions, the small circles are $SU(2)$ compositions.
    We have stretched the wires for $H$ and $T$ programs to be lines. 
    The inclined bold segments are $\textsc{cz}$ gates.}
    \label{fig:highd-comp}
\end{figure}


More generally, logical gates can be of higher depths.
Ideally, we shall employ codes with some non-transversal high-depth logical gates as symmetric matrices,
which currently we do not know of.
This is left as an interesting task for the future. 
Nevertheless, there is a method to enforce fault tolerance
by adapting the QEC dynamically during the time period of logical gates to account for the effects of the gate on the code.
The issue is also encountered for the braiding of non-Abelian anyons~\cite{NSS+08}.
Suppose a logical gate $U_L=\prod_i U_i$ as a sequence of symmetric non-logical gates $U_i$ 
(note all elementary gates and their tensor products are symmetric).
Each $U_i$ only affects a code $C$ locally,
and it defines a new code $C_i$, which is stored separately.
QEC can be performed for each $C_i$.
When the gate sequence $U_i$ is applied one after another,
it induces a sequence of codes $C_{[i]}$, which is for gates $U_{[i]}=U_i \cdots U_2 U_1$.
The support of $U_{[i]}$ will be maximal in the middle of $U_L$,
and QEC becomes challenging but can be performed in principle.

\begin{figure}[b!]
    \centering
    \includegraphics[width=0.35\textwidth]{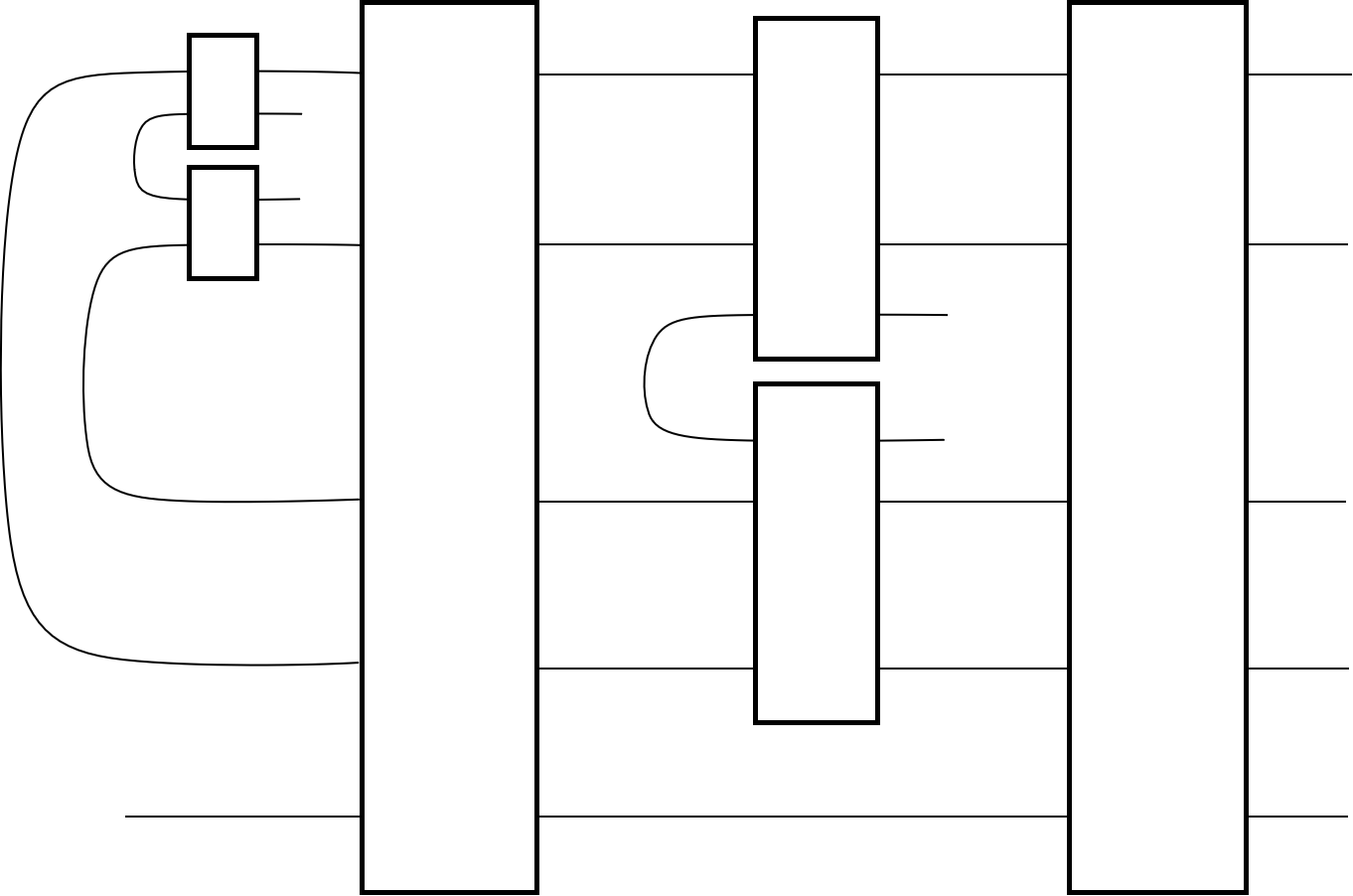}
    \caption{An example of a higher-order quantum circuit.
    The boxes are unitary gates.
    The dimension of each wire could be different.}
    \label{fig:high-comb}
\end{figure}



\section{Extensions}\label{sec:ext}

Before we conclude, we present a few direct extensions of our model. 
These are based on the key features or elements in our model, 
e.g., the usage of ebits, composition, and contraction operations. 
The extensions here can be used in our model, in principle, 
while their properties and applications remain intriguing. 
They could also be of independent interest for other tasks such as 
quantum simulation, quantum communication, 
and quantum error-correction codes.






\begin{figure}[t!]
    \centering
    \includegraphics[width=0.4\textwidth]{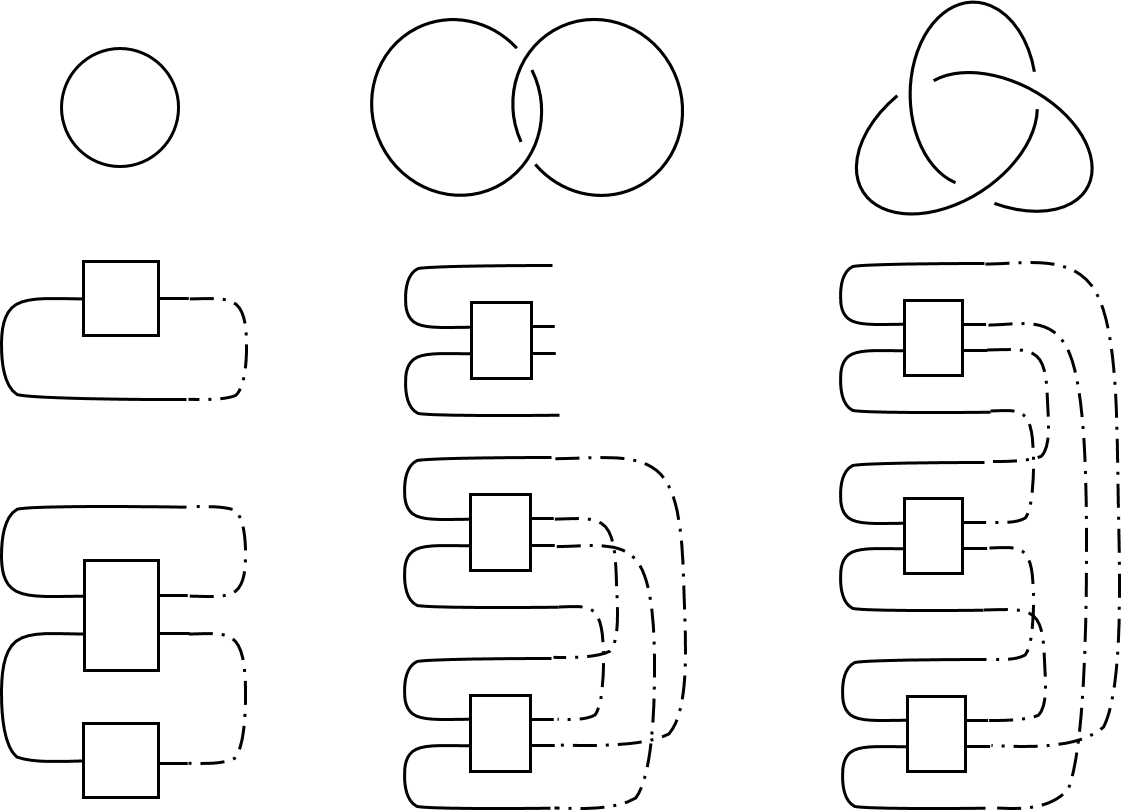}
    \caption{Primary examples of topological quantum circuits. 
    From left to right, top to bottom:
    a circle, its circuit, and a circuit for two separate circles;
    a link, the circuit for a vertex, and the circuit for the link;
    a knot, and the circuit for the knot.
    The boxes are unitary gates. The dashed lines are contractions.
    Additional qubit wires for each box are not shown explicitly.}
    \label{fig:top-circ}
\end{figure}

As for the circuit model, a quantum circuit can be imprinted on a geometric structure such as graphs or regular lattices. 
This also applies to the composition of programs, 
which can be extended from the one-dimensional flow to high-dimensional or structured flows.
For instance, we showed that two qubit-gates are connected in parallel through a $\textsc{cz}$ gate, 
which requires the $SU(2)$-covariant gate teleportation.
This can be extended to the $SU(4)$-covariant case by treating $\textsc{cz}$ and qubit gates on the equal footing.
It can lead to interesting tensor-network states, also known as PEPS.
See Fig.~\ref{fig:highd-comp} for an example.
Other Lie groups can also be used for the composition by treating the wires as representations. 
If the program states are arranged regularly, 
it can yield states with SPT orders~\cite{CGW13}.
A special class of states are valence-bond solids~\cite{AKLT87} which have global Lie-group symmetry and weak SPT order.
Indeed, if we start from empty program states, i.e., ebits,
and use the projectors onto certain representations,
this yields valence-bond solids,
which, in this setting, shall be viewed as a herald of our composition network.
It has been well established that valence-bond solids can be used for MBQC~\cite{SWP+17},
with a program served by a sequence of measurement bases, which is classical. 
On the contrary, in our scheme here the program is quantum and has been pre-stored in the ebits,
as we have discussed in section~\ref{sec:feature}.


A generalization of the tailed quantum circuits is to use higher-order quantum operations~\cite{CDP08a,CDP08,CDP09}, 
which is based on an iterative usage of the channel-state duality. 
For instance, a superchannel can also be converted into a Choi state,
and then acted upon by a channel of one higher order.
They can all be properly represented by unitary quantum circuits.
See Fig.~\ref{fig:high-comb} for an example. 
Recall that a channel is a 1-comb. 
One central difference between $n$th-order operations and $n$-combs is that 
the dimension of the former likely grows exponentially faster than the later.
That is, it is much harder to climb up the hierarchy by increasing the order than adding more input channels to a comb. 
Despite this, higher-order quantum operations lead to interesting circuits with concatenated or nested structures,
with each unitary being a ``nest.''
Nests on different levels of the hierarchy can be composed together leading to higher-order quantum combs.
These nested circuits involve nonlocal gates acting on many wires,
and may be used to describe novel quantum dynamics.

Using local gates and also the contraction operation,
we introduce another type of circuits, called topological quantum circuits,
as an extension of the tailed quantum circuits.
Intuitively, an ebit together with a contraction forms a closed loop.
A tailed quantum circuit with many tails and many contractions can be viewed as a complex of loops.
These loops are linked or knotted together due to their interactions, i.e., quantum gates.
From knot theory~\cite{Rol76}, a knot or link is formed by vertices (or crossings) and lines connecting them.
A vertex contains a top wire and a bottom wire.
Now we map a vertex to a type of circuit with two tails and two heads, 
together with possible qubit wires,
and the gate in the circuit is not fixed.
As a convention, we assign the top (bottom) wires as tails (heads).
A segment between two vertices is mapped to a contraction. 
We can see primary examples in Fig.~\ref{fig:top-circ}, 
while more complicated topological quantum circuits can also be constructed following the roles.
Without additional qubits, each diagram is an overlap between a product of Choi states
and a product of Bell states. 
If there are un-contracted qubits, a topological circuit in general prepares a multi-qubit entangled state,
without direct interactions among the qubits.
These entangled states are not apparently in the form of matrix-product states,
and their properties worth separate investigations.

\section{Discussion and conclusion}\label{sec:conc}

In this work, we present a primary model of universal quantum computer system. 
Our study extends the current formation of quantum-classical hybrid computer system,
and it shows that there are proper quantum advantages for 
information storage, processing, protection, etc.
At the meantime, a modern computer system is far more complicated than 
the original von Neumann architecture,
and this provides challenging opportunities for further development of quantum computer system.

Our study highlights the role of uncertainty principle,
revealing the fundamental difference between classical and quantum information.
The nature of quantum information is intriguing~\cite{HP20,Har04}.
Our model indicates that ebits (i.e., Bell states) are the elements for memory, 
while qubits are the elements for computation. 
On the contrary, classical bits are the elements for both classical computation and memory, 
while pbits are only used for computation.
It also appears that ebits are the analog of bits,
which is an echo of the fact that 
quantum teleportation is the analog of the one-time pad~\cite{GRT02}.
Yet Bell states are entangled and nonlocal, 
which are not the case for bits.
Furthermore, qubits are combinations of bits and pbits,
and somehow are analog signals since 
the amplitudes in superposition are carried by qubits instead of being digitized into bit strings.
In the so-called classical regime, the uncertainty principle also applies to the Fourier transform,
which is often used to analyze signals.
For the dynamics, we know that quantum computing 
generalizes permutation and stochastic operations 
to unitary operations followed by quantum measurements.
However, there is no need to convert qubits or pbits into bits,
and simulate unitary or stochastic operations by permutations.
Quantum computing is still digital in the sense that any task can be decomposed into 
elementary gates and projective measurements acting on qubits and ebits.

A key feature of our model is that the stored quantum programs are consumed after a computation. 
In order for the computer system to be reusable, 
the programs have to be restored by downloading from the internet,
if renewing the hardware of the quantum memory were not advisable.
Actually, the internet and communication network have become an indispensable part of the modern computing network.
Therefore, an extension of our model is to include a quantum internet
towards the distributed quantum computing~\cite{WEH18}. 
A major merit of distributed computing is that it distributes or breaks a computational task 
into communication among several parts, each with less requirements on its computational power. 
As a whole, it can also be viewed as a restricted QCS, e.g.,
direct operations on two QCS are not available.
This can be described by the local quantum Turing machine model~\cite{W20_Turing},
which has a close connection with matrix-product states and tensor-network states.
This points to a viable connection between tensor-network states and quantum communication.





We mentioned that we do not study QCS from the perspective of hardware,
which requests various physical and engineering tasks. 
For instance, the quantum memory in our model is a collection of quantum states.
In the sense of hardware, quantum memory refers to a hardware or device that can store quantum states,
more precisely, store the quantum systems that carry quantum information.
Devices for input and output are diverse and also require  
sophisticated information processing techniques,
but in our study they are basically bit strings generated by quantum measurements. 
QCS requires the understanding of physical problems 
including, but not limited to, dissipation, efficiency, energy cost, 
role of fundamental laws, mechanism of almost all devices, etc. 
More broadly, as a giant artificial quantum system, 
QCS is a subject to reveal the 
interplay between quantum physics and modern technologies based on
control, system, and information theory.



\section*{Acknowledgements}

This work has been supported by the National Natural Science Foundation of China 
(Grant No. 12047503 \& No. 12105343).
Previous conversations with I. Affleck, P. Hayden, R. Laflamme, H. Nautrup, T. Shi, Y. Wang, J. Watrous, Y. Wu, Y. Yang, S. Yi, and G. Zhu are acknowledged. 

\bibliography{ext}{}
\bibliographystyle{elsarticle-num-names}

\end{document}